\documentstyle[12pt,epsf]{article}

\newcommand{\be}{\begin{equation}}
\newcommand{\ee}{\end{equation}}
\newcommand{\nn}{\nonumber}
\newcommand{\beba}{\begin{equation}\begin{array}{lcl}}
\newcommand{\eaee}{\end{array}\end{equation}}
\newcommand{\bea}{\begin{eqnarray}}
\newcommand{\eea}{\end{eqnarray}}
\newcommand{\ba}{\begin{array}}
\newcommand{\ea}{\end{array}}

\newcommand{\ns}{\normalsize}
\newcommand{\refs}[1]{(\ref{#1})}
\def\bal{{\mbox{\boldmath $\alpha$}}}
\def\bla{{\mbox{\boldmath $\lambda$}}}
\def\bt{{\mbox{\boldmath $\tau$}}}
\def\bq{{\bf q}}
\def\a{\alpha}
\def\b{\beta}
\def\g{\gamma}

\def\d{\delta}
\def\e{\epsilon}

\def\f{\phi}

\def\m{\mu}
\def\n{\nu}

\def\r{\rho}

\def\t{\tau}

\def\D{\Delta}

\def\L{\Lambda}

\begin{document}
\begin{titlepage}
\title{{\large\bf Cosmological Solutions of Type II
        String Theory}\setcounter{footnote}{0}\thanks{Work supported in
        part by DOE under Contract No.~DOE--AC02--76--ERO--3071.}\\
                          \vspace{-4cm}
                          \hfill{\ns UPR-711T\\}
                          \hfill{\ns hep-th/9608195\\[.1cm]}
                          \hfill{\ns August 1996}
                        \vspace{2.3cm} }
\author{Andr\'e
        Lukas\setcounter{footnote}{6}\thanks{Supported by Deutsche
        Forschungsgemeinschaft (DFG) and
        Nato Collaborative Research Grant CRG.~940784.}~~,
        Burt A.~Ovrut and Daniel Waldram\\[0.5cm]
        {\ns Department of Physics, University of Pennsylvania} \\
        {\ns Philadelphia, PA 19104--6396, USA}\\}
\date{}
\maketitle
\begin{abstract} \baselineskip=6mm
We study cosmological solutions of type II string theory with a metric
of the Kaluza--Klein type and nontrivial Ramond--Ramond forms. It is
shown that models with only one form excited can be integrated in general. 
Moreover, some interesting cases with two nontrivial forms can be solved 
completely since they correspond to Toda models. We find two types of 
solutions corresponding to a negative time superinflating phase and a 
positive time subluminal expanding phase. The two branches are separated by 
a curvature singularity. Within each branch the effect of the forms is to 
interpolate between different solutions of pure Kaluza--Klein theory.
\end{abstract}
\thispagestyle{empty}
\end{titlepage}
Presently, one of the main challenges in string theory is to find solutions
compatible with the standard picture of early-universe cosmology. Though
successful in reproducing most generic features of low-energy particle
phenomenology string theory poses serious problems in realizing even
basic ingredients of cosmology such as inflation.

So far, most studies of string cosmology were concerned with solutions of
the heterotic string effective action~\cite{heter_sol}. The basic picture
is provided by the classical ``rolling--radii''--solutions of
Mueller~\cite{mueller,roll2} which are Kaluza--Klein--type~\cite{kk}
solutions
with time-dependent dilaton and radii of the internal and external space.
Classical solutions with a nontrivial Neveu--Schwartz (NS) 2--form
have also been found~\cite{NS_sol} and the $O(d,d)$ symmetry of the
low energy effective action proves to be useful in their
classification~\cite{mei_ven}.

With the discovery of string dualities~\cite{s_dual} the general picture of 
string theory has changed dramatically. The five formerly unrelated 
consistent string theories are now believed to correspond to certain
limiting cases of one underlying theory, called M--theory whose effective
low-energy Lagrangian is given by 11--dimensional supergravity~\cite{duff}.
Correspondingly, 11-dimensional and type II supergravities may be of
direct importance for particle phenomenology as well as
cosmology~\cite{dine} and a study of cosmological solutions in these
theories appears to be an important issue.

A large amount of work has been devoted to the study of classical solutions
in type II theories, however almost all of this work is concentrated on
black hole and membrane solutions~\cite{duff_rep}. The characteristic property
of these solutions which have played a major r\^ole in uncovering string
dualities is that they are charged with respect to the Ramond--Ramond (RR)
forms of type II string theory.

\vspace{0.4cm}

In this letter we present a first study of cosmological solutions in
type II theories with nontrivial RR--forms. The main ideas will be
illustrated by a few examples. A systematic study of this class of
cosmological solutions will be presented elsewhere~\cite{long_paper}.

We consider a $D$--dimensional space time (with $D=10$ in most examples)
of Kaluza--Klein type split up into a number of maximally symmetric flat
subspaces of dimensions $d_i$ with scale factors $a_i=\exp (\a_i)$. This
choice along with a time dependent dilaton $\f$ appears to be the simplest
appropriate to cosmology and it is
remarkably close to the space--times corresponding to membrane solutions.
The r\^ole of the coordinates transverse to the membrane is here played by
time and in contrast to real membranes the ``worldsheet'' is purely
spacelike. Also the two types of membrane solutions -- elementary and
solitonic -- find their natural analog in our cosmological setting~: 
the symmetry of the Kaluza--Klein space is compatible with two types of field
strengths one having a nontrivial time--direction (elementary) the other
one being nonvanishing in spatial directions only (solitonic).

As we will see the forms provide an effective potential for the scale
factors $\a_i$ and the dilaton which depends on the type of forms
that are considered and the specific Ansatz.
It generates interesting dynamics of the model beyond what is known
for the pure Kaluza--Klein case~\cite{mueller}. Fortunately, general
solutions can be found as long as only one form is turned on. This will
be demonstrated below for a type IIB example with a nontrivial RR 2--form.
For more than one form this is no longer generally possible. However,
as we will show, a number
of interesting cases correspond to Toda--models~\cite{toda,kostant} which can
be integrated completely. The use of Toda--theory in finding Kaluza--Klein
dyon solutions has been advocated in ref.~\cite{gib_lee} and it has first been
applied to cosmological models in ref.~\cite{gib_mae}. Applications to
cosmologies with perfect fluids can be found in ref.~\cite{russian_guys}.
Recently, also nonextremal soliton solutions in supergravity have been
studied using Toda theory~\cite{lu_pope}.

We will present examples within IIA supergravity having an elementary
RR 3--form and a solitonic RR 1--form or an elementary NS 2--form) leading 
to a $SU(2)^2$  or $SU(3)$  Toda theory respectively.

\vspace{0.4cm}

The following general picture emerges form the study of these solutions.
In all cases we find two types of solutions corresponding to a
negative ($-$) and positive ($+$) time branch.  Except for specific initial
conditions these two branches are separated from each other by a
curvature singularity. For both branches we can find certain limiting
cases in comoving time $t$ (such as $t\rightarrow\pm\infty$, $t$ close to
the singularity or in some intermediate range) where the forms are
effectively turned off and the theory behaves like a pure Kaluza--Klein
theory with a time-dependent dilaton, i.~e.~it is described by Hubble
parameters
\bea
 H_i &=& \dot{\a}_i\simeq \frac{p_i}{t} \nn \\
 H_\f &=& \dot{\f}\simeq \frac{p_\f}{t} \label{Hubble}
\eea
satisfying the constraints
\bea
 \sum_{i=1}^{n}d_ip_i &=& 1 \nn \\
 \frac{1}{2}p_\f^2+\sum_{i=1}^{n}d_ip_i^2 &=& 1 \; . \label{constraints}
\eea
In particular, these constraints imply that $|p_i| < 1$ always. For the
($+$) branch ($t>0$) this results in a subluminal, ``radiation like''
expansion ($p_i$ positive) or contraction ($p_i$ negative). For the
($-$) branch ($t<0$) the space will be superinflating for $p_i<0$ (the
horizon shrinks) or collapsing for $p_i>0$. Though the properties of the
($-$) branch are very similar to the superinflating phase of pre--big--bang
models~\cite{pbb} they arise in the Einstein frame as opposed to the
string frame which is used in those models.

The forms become operative during the transition periods between the above
asymptotic regions. Their effect can be described by a mapping
$p_i\rightarrow \tilde{p}_i$ changing the expansion coefficients $p_i$ in one
asymptotic region to $\tilde{p}_i$ in another asymptotic region.  For example
a universe in the ($+$) branch, split such that $(d_1,d_2)=(3,6)$, with a
contracting 3--dimensional and an expanding 6--dimensional subspace at
early times $t$, can be turned into the ``desired'' one with expanding
3--dimensional and contracting 6--dimensional subspace at late times $t$.

\vspace{0.4cm}

Let us now be more specific and discuss some explicit examples. The action
in the Einstein frame is of the general form
\be
 S=\int d^Dx\;\sqrt{-g}\left[ R-\frac{4}{D-2}(\partial\f )^2-\sum_r
    \frac{1}{2(\d_r+1)!}e^{-p_r\f}F_r^2\right] \label{action}
\ee
with the $D$--dimensional metric $g_{MN}$, the dilaton $\f$ and a number
of form fields $F_r=dA_r$, $r=1...m$ of degree $\d_r$. For type II
theories the dilaton couplings $p_r$ are given by
\be
 p_r = \left\{ \ba{cll} \frac{8}{D-2}&{\rm NS}&{\rm 2-form}\\
                      \frac{4\d_r -2(D-2)}{D-2}&{\rm RR}&\d_r{\rm -form}
             \ea\right.\; .
\ee
Examples in $D=11$ supergravity can be accommodated by setting $\f =$ const.
In eq.~\refs{action} we have neglected Chern--Simons terms for simplicity
and our examples will be chosen such that this is consistent. 
Other solutions with nonvanishing Chern--Simons terms can presumably be
generated from ours via duality~\cite{bergs}.

\vspace{0.4cm}

For our first example we will use the simplest split of space
$(d_1,d_2)=(3,6)$ and a solitonic type IIB RR 2--form $F$ in the $d_1=3$
subspace. Correspondingly, the Ansatz for the metric, the 2--form and the
dilaton reads
\bea
 ds^2 &=& -N^2(\t )d\t^2+e^{2\a_1}\sum_{\m=1}^{3}dx_\m^2+e^{2\a_2}
           \sum_{m=4}^{9}dx_m^2\nn\\
 F_{\m\n\r} &=& e^{-6\a_1} u\, \e_{\m\n\r}\\
 \f &=& \f (\t )\nn\; ,
\eea 
where $u$ is a constant~\footnote{For the totally antisymmetric tensor
we use the convention $\e^{123}=1$ and
$\e_{\m\n\r}=g_{\m\a}g_{\n\b}g_{\r\g}\e^{\a\b\g}$.}. Using the gauge
$N=\exp (3\a_1-6\a_2-\f )$ we arrive at the following equations of motion
\bea
 \frac{d}{d\t}\left( e^{12\a_2+\f}(\a_1'+3\a_2')\right) &=& 0\nn\\
 \frac{d}{d\t}\left( e^{12\a_2+\f}(3\a_1'+5\a_2')\right)-\frac{u^2}{2} &=&
  0\nn\\
 \frac{d}{d\t}\left( e^{12\a_2+\f}\f '\right) +\frac{u^2}{2} &=& 0 \\
  e^{12\a_2+\f}(6{\a_1'}^2+36\a_1'\a_2'+30{\a_2'}^2-\frac{1}{2}{\f '}^2)-
  \frac{u^2}{2}&=& 0\nn
\eea
for $\a_1$, $\a_2$, $\f =\a_\f$ and $N$, respectively. The prime denotes the
derivative with respect to $\t$. Their general solution is given by
\be
 \a_I = c_I\ln (\t_1-\t )+w_I\ln\left(\frac{\t}{\t_1-\t}\right)
          +k_I \label{sol1} \\
\ee
where the index $I=(i,\f )$ labels the scale factors and the dilaton.
The integration constants $w_I$ and $k_I$ are subject to the constraints
\bea
 12w_2+w_3 &=& 1 \nn\\
 12w_1^2+72w_1w_2+60w_2^2 &=& w_3^2 \label{cons1} \\
 \exp (12k_2+k_3) &=& u^2 \nn \; .
\eea
The range of the time parameter $\t$ is specified by $0 < \t < \t_1$.
It is remarkable that the numerical coefficients
\be
 (c_1,c_2,c_\f )=\frac{1}{8}(-3,1,4)
\ee
in front of the first term of the solution~\refs{sol1} coincide with
those of a solitonic 5--brane solution~\cite{duff_rep}. In fact, this first
term represents the analog of such a solution, which is characterized by
the proportionality of $\a_1$, $\a_2$ and $\f$. This raises the immediate
question of whether cosmological BPS solutions which preserve one half of
the supersymmetries can be found. In the case at hand, as well as in
the other 10--dimensional examples with one form excited, this turns out to be
impossible since the second term in eq.~\refs{sol1} cannot be consistently
set to zero. In fact, inspection of the supersymmetry transformation
shows that this is precisely what is required for a supersymmetric solution.
Still, the form of eq.~\refs{sol1} is so tantalizingly close
to allowing a BPS solution that one might hope for such a possibility in
related models.

To discuss the cosmological properties of this model it is useful to
have expressions for the gauge parameter $N$ and the curvature~:
\bea
 N &\sim& (\t_1 -\t )^{-x-\D -1}\,\t^{x-1} \\
 R &\sim& (\t_1 -\t )^{2(x+\D )}\,\t^{-2x} P(\t )\label{R}\; .
\eea
The quantities $x$, $\D$ are given by
\bea
 x &=& 3w_1+6w_2 \nn\\
 \D &=& \frac{3}{8}
\eea
and $P$ denotes a second-order polynomial in $\t$. We remark that the
comoving time $t$ can be explicitly expressed in terms of $\t$ by
integrating $dt=N(\t )d\t$ leading to hypergeometric functions.
The constraints~\refs{cons1} admit two ranges for the parameter $x$,
$x >0$ and $x < - \D$. This implies the following mapping of ranges
\be 
 \t\in \left[ 0,\t_1 \right]\;\rightarrow\; t\in\left\{ \ba{llll} 
                \left] -\infty ,t_1\right]&{\rm for}&x<-\D\; ,&
                -\;{\rm branch}\\
                \left[ t_0,+\infty \right[ &{\rm for}&x>0\; ,&+\; {\rm branch}
                \ea\right.
\ee
between $\t$ and the comoving time $t$. Since the intermediate region
$-\D < x <0$ is forbidden there is no way to connect the two branches
within this class of models. The above expression for $R$ shows that
generically there will be a curvature singularities at the finite ends of
both branches. For special initial conditions, however, the polynomial $P$
can partially cancel the prefactors in eq.~\refs{R} and the curvature
singularity disappears.
This occurs for the ($-$) branch if $w_3=c_3$ and $x\geq -\D -1/2$ and for
the ($+$) branch if $w_3=0$ and $x\leq 1/2$. Such a cancellation is very
similar to what happens in the singularity-free model of
ref.~\cite{kou_lust} which was derived from a WZW--model.
\centerline{\epsfbox{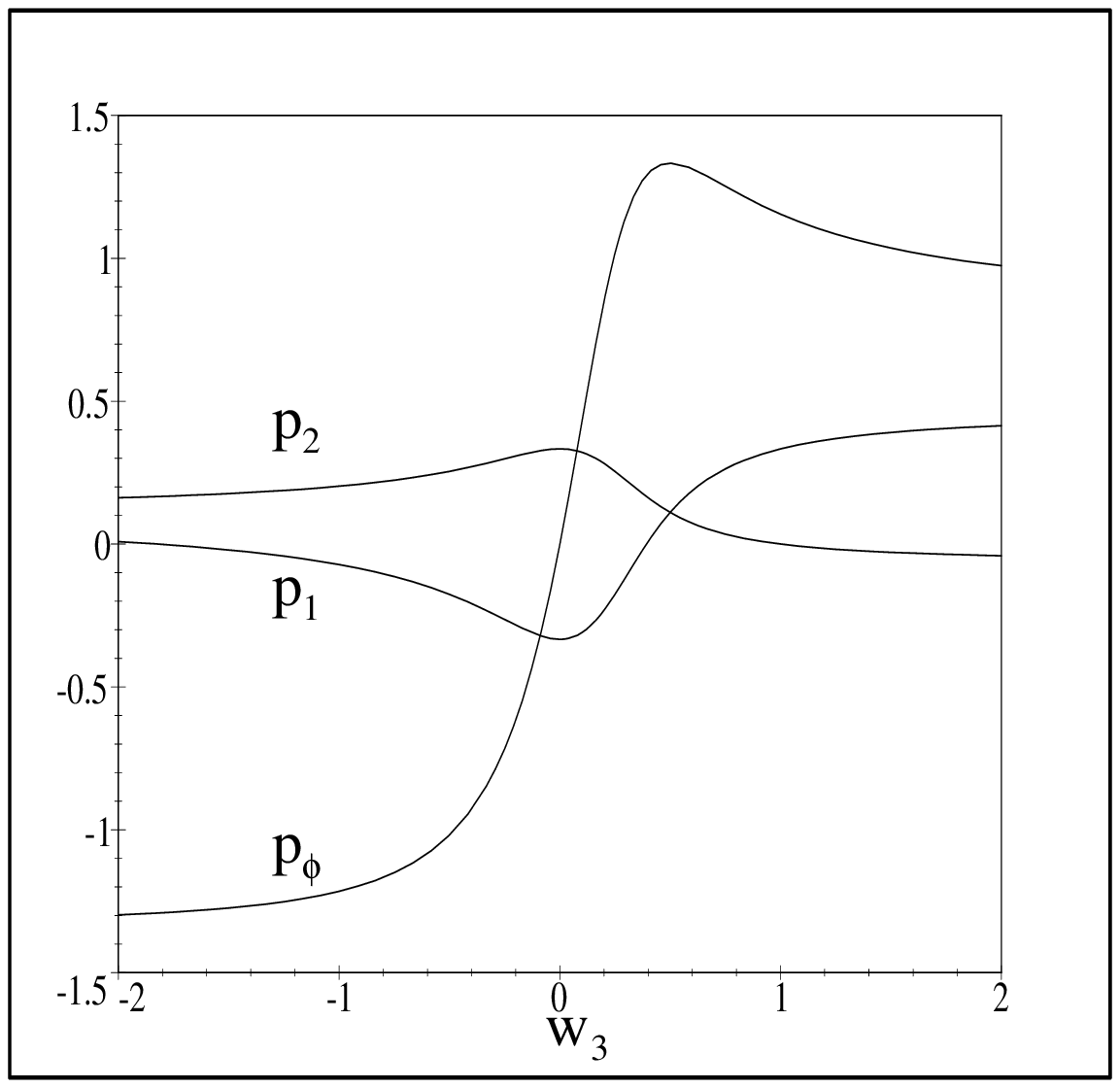}}
\centerline{\em Fig 1: Expansion coefficients for the
            ($+$) branch at $t\simeq t_0$.}
\vskip 0.4cm
The asymptotic regions where the form is effectively turned off are specified
by $\t\simeq 0$ and $\t\simeq\t_1$ and the corresponding expansion
coefficients in eq.~\refs{Hubble} read
\bea
 \t\simeq 0 &:& p_I^{(0)} = \frac{w_I}{x} \nn \\
 \t\simeq \t_1 &:& p_I^{(1)} = \frac{w_I-c_I}{x+\D}\; .
\eea
After using the constraints~\refs{cons1} they still depend
on one free parameter, which here we take to be $w_3$. An example of this 
dependence for the ($+$) branch
is given in fig.~1 and fig.~2. It can be seen that the
3--dimensional subspace which for small $w_3$ is contracting at early
times $t\simeq t_0$ is turned to expansion at late time
$t\rightarrow\infty$.\\
\centerline{\epsfbox{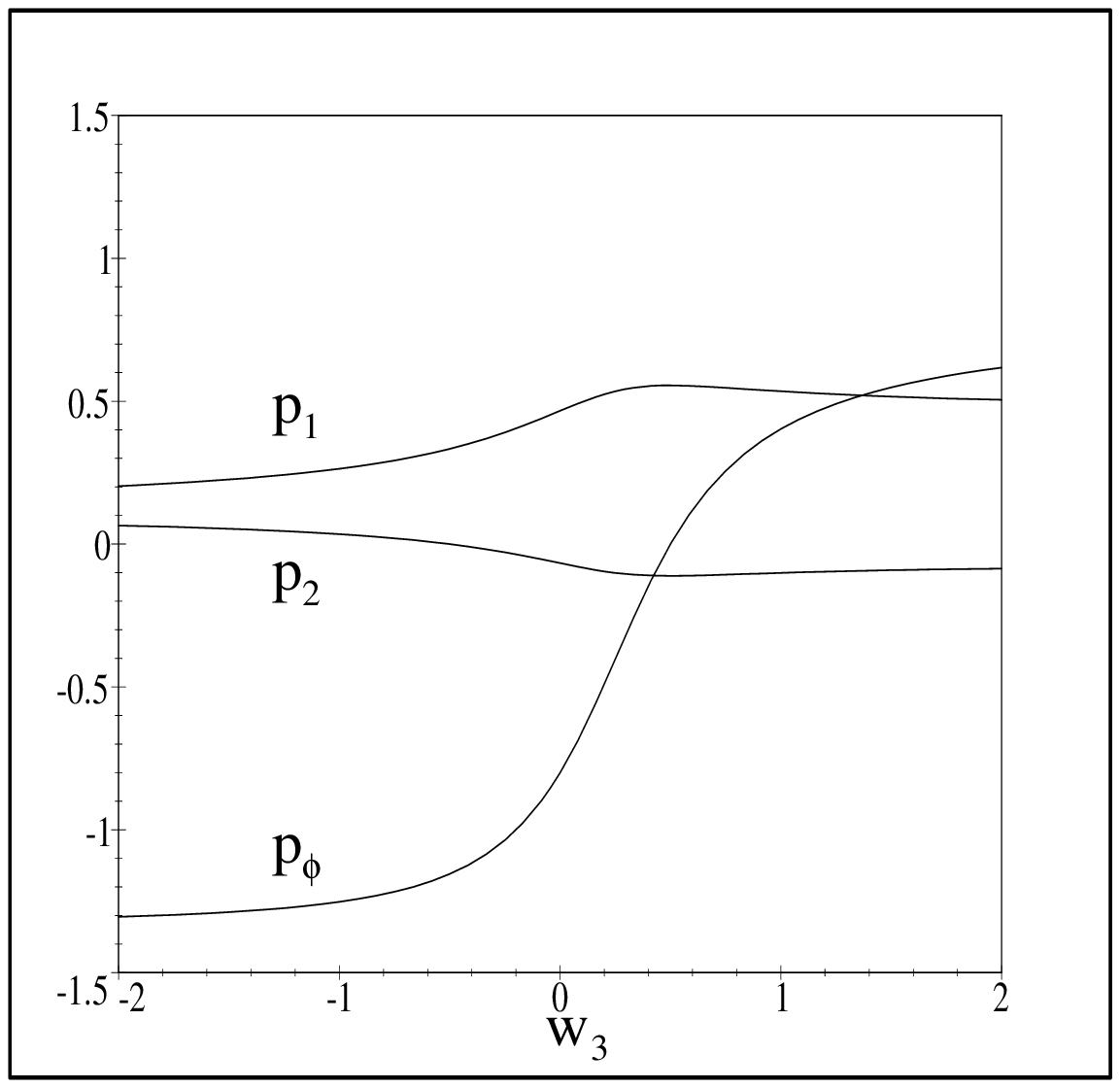}}
\centerline{\em Fig 2: Expansion coefficients for the ($+$) branch at
            $t\rightarrow\infty$.}
\vskip 0.4cm
The converse is
true for the 6--dimensional ``internal'' space. In the ($-$) branch
an early $t\rightarrow -\infty$ contraction of the 3--dimensional subspace
and a hyperinflating expansion of the 6--dimensional subspace can similarly be
turned into its converse as $t$ approaches $t_1$. Though this
``desirable'' situation is clearly arranged in the present context
by choosing initial conditions in the correct range it is by no means
obvious that this was possible at all. In a very similar model with
a $(d_1,d_2)=(3,6)$ or $(d_1,d_2)=(3,7)$ split but an {\em elementary} 3--form
in the $d_1=3$ subspace, to be realized in the context of type IIA or
11--dimensional supergravity, it is impossible to have 3 expanding and
6 or 7 contracting dimensions as $t\rightarrow\infty$.

\vspace{0.4cm}

Next, we would like to study how the above picture generalizes once two
forms are turned on. We choose a split $(d_1,d_2,d_3)=(3,2,4)$, with an
elementary IIA 3--form $F^{(3)}$ in the $d_1=3$ subspace and a solitonic IIA
1--form $F^{(1)}$ in the $d_2=2$ subspace. This results in the following
Ansatz~:
\bea
 ds^2 &=& -N^2(\t )d\t^2 +e^{2\a_1}\sum_{\m =1}^{3}dx_\m^2+
           e^{2\a_2}\sum_{m =4}^{5}dx_\m^2+e^{2\a_3}\sum_{a =6}^{9}dx_\m^2\nn\\
 F^{(3)}_{0\m\n\r} &=& e^{-6\a_1}h_1'\;\e_{\m\n\r} \nn \\
 F^{(1)}_{mn} &=& e^{-4\a_2}v_2\;\e_{mn} \label{ansatz2}\\
 \f &=& \f (\t )\; .\nn
\eea
The model simplifies considerably in the harmonic gauge
$N=\exp (3\a_1+2\a_2+4\a_3)$ where it corresponds to a Hamiltonian system.
The equation of motion for $F^{(3)}$ can be integrated to give
$h_1'=v_1\exp (6\a_1-\f /2)$ with a constant $v_1$. This result can be used
to replace $h_1'$ in the other equations. To simplify
the notation it is useful to introduce the vector $\bal = (\a_I)=(\a_i,\f )$
and the metric $G_{IJ}$ with $G_{ij}=2(d_i\d_{ij}-d_id_j)$,
$G_{i\f}=G_{\f i}=0$, $G_{\f\f}=8/(D-2)$. The equations of motion for $\bal$
can be obtained by a variation of the Lagrangian
\bea
 {\cal L} &=& \frac{1}{2}{\bal '}^TG\bal ' - U \nn\\
 U &=& \frac{v_1^2}{2}\exp (\bq_1.\bal )+\frac{v_2^2}{2}\exp (\bq_2.\bal )
 \label{lag}
\eea
The influence of the forms on the effective potential $U$ is encoded in
the vectors $\bq_1 = (6,0,0,-1/2)$ and $\bq_2 = (6,0,8,3/2)$ for the
3-- and the 1--form respectively. The above system has to be supplemented
by the Hamiltonian constraint
\be
 {\cal H} = \frac{1}{2}{\bal '}^TG\bal ' + U = 0\; \label{ham}.
\ee
Contact with Toda theory can be made if the matrix $<\bq_r,\bq_s>$
computed with the scalar product defined by
$<{\bf x},{\bf y}>={\bf x}^TG^{-1}{\bf y}$ is proportional to the Cartan
matrix of a semi--simple Lie group. In fact, in our case we have
$<\bq_1,\bq_1>=<\bq_2,\bq_2>=4$ and $<\bq_1,\bq_2>=0$ which corresponds to
the Cartan matrix of $SU(2)^2$. This allows one to decouple the two
exponentials in the potential $U$ and to find the explicit
solution~\cite{russian_guys}
\be
 \bal = \sum_{r=0}^{3}\r_rG^{-1}\bq_r \label{sol2}
\ee
with $\bq_1$, $\bq_2$ as above, $\bq_0 = (3,1,-8,0)$, $\bq_3 = (4,0,8,1)$
and the functions
\bea
 \r_r &=& -\frac{1}{4}\ln\left( \frac{2v_r^2}{k_r^2}
         \cosh^2 (|k_r|(\t -\t_r))\right)\; ,\quad r=1,2 \label{sol2_12} \\
 \r_r &=& k_r(\t -\t_r )\; ,\quad r=0,3 \; . \label{sol2_03}
\eea
The time parameter $\t$ ranges over the full real axis
and $k_r$, $\t_r$, $r=0...3$ are integration constants.
The Hamiltonian constraint~\refs{ham} turns into
\be
 k_1^2+k_2^2+\frac{8}{3}k_3^2=\frac{1}{4}k_0^2\; \label{cons2}.
\ee
As in the first example this condition can be used to prove the existence
of a ($+$) and a ($-$) branch with general properties as discussed in the
introduction. The two Eigenmodes $\r_1$ and $\r_2$ describe the effect of
the 3-- and the 1--form, respectively. The corresponding integration constants
$\t_1$ and $\t_2$ have a very simple interpretation as can be seen from
eq.~\refs{sol2_12}. The 3--form is operative around $\t\simeq\t_1$, the
1--form around $\t\simeq\t_2$. This allows for three asymptotic Kaluza--Klein
regions $\t\rightarrow -\infty$, $\t_1\ll\t\ll\t_2$ (or $\t_2\ll\t\ll\t_1$)
and $\t\rightarrow +\infty$ where the forms are effectively turned off.
The two forms act independently, each being responsible for one of the
two transitions. In this sense, the model can be interpreted as a time
sequence of two simple models each with just one form turned on.

\vspace{0.4cm}

Finally, we would like to discuss an example which does not lead to a
``decoupling'' of the two forms in the above sense. We start with the
same Ansatz~\refs{ansatz2} as in the previous example except for the
solitonic RR 1--form $F^{(1)}$ which we replace by
an elementary NS 2--form $F^{(2)}$ in the same $d_2=2$ subspace and with
the Ansatz
\be
 F^{(2)}_{0mn} = e^{-4\a_2}h_2'\, \e_{mn}\; .
\ee h
The equation of motion for $F^{(2)}$ is used to replace $h_2'$ by
$h_2'=v_2\exp (4\a_2+\f)$ and in the harmonic gauge we end up with the
same system specified by~\refs{lag} and~\refs{ham} but with $\bq_2$ replaced
by $\bq_2 = (0,4,0,1)$. Now the scalar products turn out to be
$<\bq_1,\bq_1>=<\bq_2,\bq_2>=4$ and $<\bq_1,\bq_2>=-2$ indicating that we
are dealing with an $SU(3)$ Toda theory.
The solution~\cite{kostant,russian_guys} takes the form~\refs{sol2} with
$\r_0$ and $\r_3$ given by eq.~\refs{sol2_03} and $\bq_0 = (1,2/3,1,0)$,
$\bq_3 = (0,-4/3,0,1)$. For $\r_1$ and $\r_2$ we get
\be
 \r_r = -\frac{1}{2}\ln\left(\sum_{\bla\in\L_r}b_r(\bla )
        \exp (\bla .{\bf k}\, \t - \bla .\bt)\right)\; ,\quad r=1,2\; ,
 \label{sol3}
\ee
where $\L_r$ are the weight systems of the two fundamental $SU(3)$
representations, namely $\L_1 = \{ (1,0),(-1,1),(0,-1)\}$ for ${\bf 3}$ and
$\L_1 = \{ (0,1),(1,-1),$ $(-1,0)\}$ for ${\bf\bar{3}}$ and
${\bf k} = (k_1,k_2)$, $\bt = (\t_1,\t_2)$. The coefficients $b_r(\bla )$ can
be expressed as
\be
 \ba{lllllll}
  b_1(1,0) &=& 2v_1^2\,\frac{2k_2-k_1}{D}&
  &b_2(0,1) &=& 2v_1^2\,\frac{2k_1-k_2}{D}\\
  b_1(-1,1) &=& 2v_1^2\,\frac{k_1-k_2}{D}&
  &b_2(1,-1) &=&2v_2^2\,\frac{k_1-k_2}{D}\\
  b_1(0,-1) &=& 2v_2^2\,\frac{2k_1-k_2}{D}&
  &b_2(-1,0) &=& 2v_2^2\,\frac{2k_2-k_1}{D}
 \ea
\ee
with $D=(k_1-2k_2)(k_1-k_2)(2k_1-k_2)$. The Hamiltonian constraint now reads
\be
 k_1^2-k_1k_2+k_2^2+\frac{4}{3}k_3^2=\frac{1}{24}k_0^2\;
\ee
and we demand  $2k_1-k_2>0$, $2k_2-k_1>0$ (${\bf k}$ is in the Weyl chamber)
to ensure a positive argument of the logarithm in eq.~\refs{sol3}.
The picture of the previous example applies in this case as well.
For $\t\rightarrow\pm\infty$ we have two asymptotic Kaluza--Klein
regions and depending on the choice of integration constants two
intermediate Kaluza--Klein regions can occur.

\vspace{0.4cm}

In conclusion, we have presented a first study of cosmological solutions
with nontrivial RR--fields in type II string theory. We have shown that
models with one RR--form are generally solvable and a number of
interesting examples with two forms can be solved because they
correspond to Toda models. It turns out that the potential provided by
the RR--forms generates interesting dynamics which
allows interpolation between different pure Kaluza--Klein
states. A key ingredient to a direct cosmological application of our results
is certainly to connect the hyperinflating and the subluminal branch which
we found in all our solutions.
This turned out to be impossible within the class of models we have
analyzed. It is likely that there is no classical way at all to achieve
such a connection and that nonperturbative physics has to be invoked for that.
As for pre--big--bang models which suffer from the same problem one
might speculate that a duality map between the branches corresponds to
the correct transition mechanism.

Our Ansatz and the solution~\refs{sol1} turned out to be very
close to membrane solutions which might suggest the possibility of
a cosmological BPS state preserving half of the supersymmetries. Such
a solution would certainly have very interesting applications to
density fluctuations in the early universe which presumably could be
``counted'' in a similar way as the states of some extremal
black holes~\cite{bh}.

\vspace{0.4cm}

{\bf Acknowledgment} A.~L.~is supported by a fellowship from
Deutsche Forschungsgemeinschaft (DFG). 
\end{document}